\documentclass[12pt]{article}

\usepackage{scicite}

\usepackage{times}

\usepackage{amssymb}
\usepackage{amsmath}
\usepackage{graphicx}
\usepackage[misc]{ifsym}
\usepackage{mathptmx}      
\usepackage{cite}
\usepackage{color}
\usepackage{epsfig}
\usepackage{epstopdf}

\newenvironment{sciabstract}{%
\begin{quote} \bf}
{\end{quote}}

\title{How the reversible change of contact network affects the epidemic spreading}

\author
{Xincheng Shu,$^{1,2}$ Zhongyuan Ruan,$^{1\ast}$\\
\\
\normalsize{$^{1}$Institute of Cyberspace Security, Zhejiang University of Technology,  Hangzhou,  310023, China}\\
\normalsize{$^{2}$Department of Electrical Engineering, City University of Hong Kong, Hong Kong, 999077, China}\\
\\
\normalsize{$^\ast$ \textbf{Correspondence}: zyruan@zjut.edu.cn}
}

\date{}

\begin{document} 


\baselineskip24pt


\maketitle


\begin{sciabstract}
  The mobility patterns of individuals in China during the early outbreak of the COVID-19 pandemic exhibit reversible changes --- in many regions, the mobility first decreased significantly and later restored. Based on this observation, here we study the classical SIR model on a particular type of time-varying network where the links undergo a freeze-recovery process. We first focus on an isolated network and find that the recovery mechanism could lead to the resurgence of an epidemic. The influence of link freezing on epidemic dynamics is subtle. In particular, we show that there is an optimal value of the freezing rate for links which corresponds to the lowest prevalence of the epidemic. This result challenges our conventional idea that stricter prevention measures (corresponding to a larger freezing rate) could always have a better inhibitory effect on epidemic spreading. We further investigate an open system where a small fraction of nodes in the network may acquire the disease from the ``environment" (the outside infected nodes). In this case, the second wave would appear even if the number of infected nodes has declined to zero, which can not be explained by the isolated network model.
\end{sciabstract}


\section*{Introduction}

\begin{figure}
\epsfig{figure=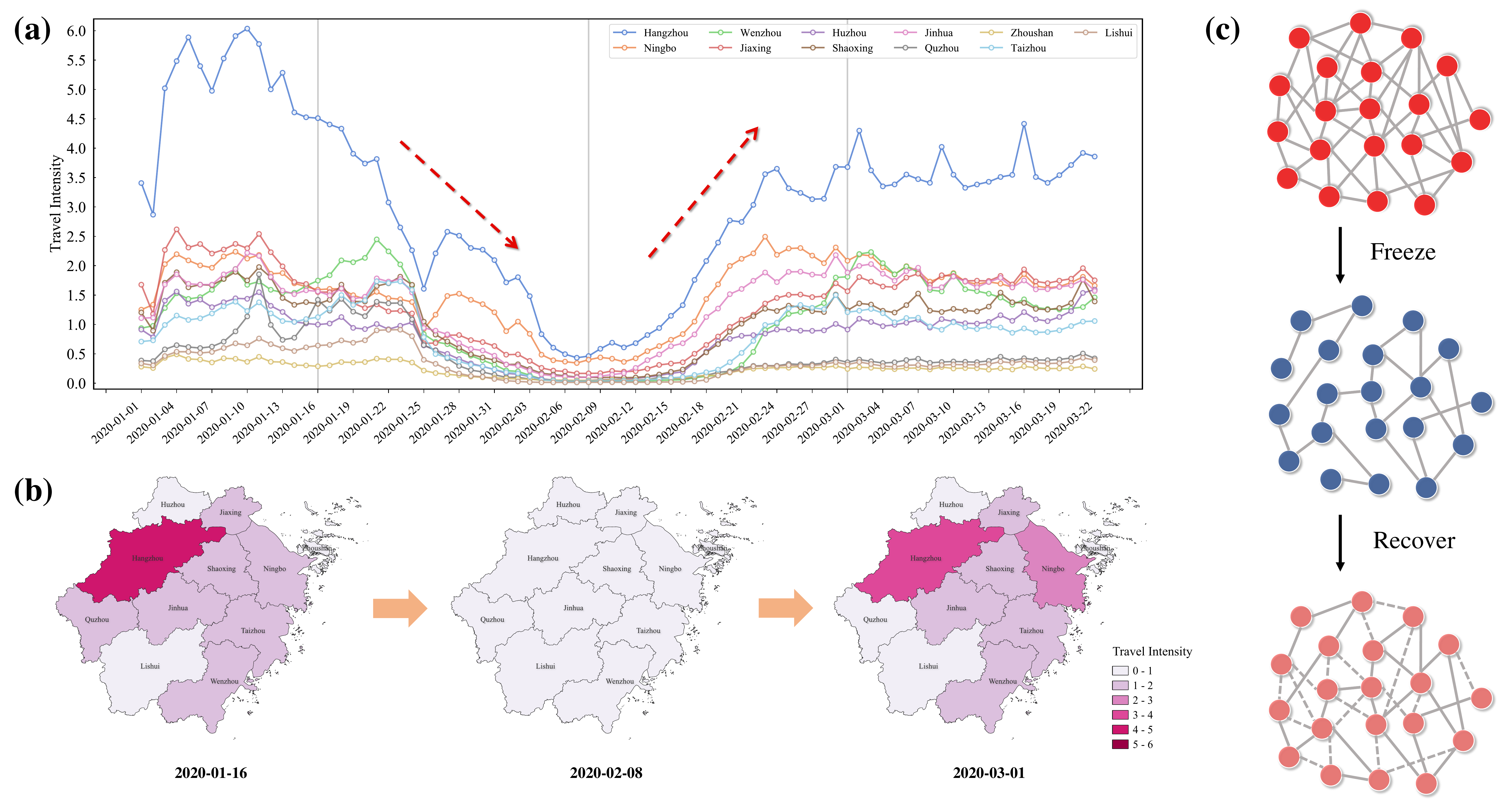,width=1.0\linewidth} \caption{(color online). (a) Evolution of the intracity travel intensity for all $11$ cities in Zhejiang province in China from $1$ January to $22$ March, $2020$. (b) Visualization of the intracity travel intensity for each city in Zhejiang province on three typical dates. (c) Reversible change in the contact network. The travel restriction during the COVID-19 disease outbreak led to many people cutting their physical connections to others over a period of time. After that, these links started to recover, becoming active again. The intracity travel intensity data are collected from Baidu Map Smart Eye.} \label{Fig:1}
\end{figure}

During the early prevalence of COVID-19 (coronavirus disease 2019) in mainland China, the Chinese government implemented a series of containment policies aiming to mitigate the spread of the disease \cite{Maier:2020,Zhang1:2020,Zhang2:2020,Schlosser:2020,Ventura:2022}. These policies led to substantial changes in human mobility patterns \cite{Gibbs:2020,Badr:2020} [see Fig. \ref{Fig:1} (a) and (b)]. Figure \ref{Fig:1} (a) shows the intracity travel intensity (defined as the ratio of the number of individuals traveling in the city to the number of people living in it \cite{Yang:2022}) for each city in Zhejiang province in China as a function of time from $1$ January to $22$ March, $2020$. We observe that mobility was considerably reduced at first, and after a turning point (around $8$ February, $2020$), it started to restore and finally returned to the pre-pandemic state, displaying a clear reversible process. Since mobility directly affects the average number of contacts per person, it is expected that the contact patterns would change correspondingly: we assume that the links (connections) between individuals would experience a freeze-recovery process (the frozen links are equivalent to being removed from the contact network temporarily), as shown in Fig. \ref{Fig:1} (c). 

Many studies have focused on how the time-varying contact networks may affect the epidemic dynamics \cite{Meloni:2009,Ruan:2012,Yang:2019,Gross:2006,Marceau:2010,Yang:2012,Karsai:2011,Perra:2012,Holme:2012,Pozzana:2017,Zino:2016,Zanette:2008,Vazquez:2007}. For instance, it has been shown that adaptive networks (individuals avoid contact with the infected) could give rise to rich dynamics like hysteresis and first-order transitions \cite{Gross:2006}. Apart from this, some research focused on the situation in which the contact structures evolve independently of the dynamical process \cite{Karsai:2011,Perra:2012,Holme:2012,Pozzana:2017,Zino:2016,Vazquez:2007}. For example, it is demonstrated that temporal heterogeneities in contacts can slow down spreading \cite{Karsai:2011}. Besides, a notable finding shows that in the activity-driven networks, the epidemic threshold does not depend on the time-aggregated network representation, but is a function of the interaction rate of the nodes \cite{Perra:2012}. 

Despite these progresses, the recovery process of the contact networks has not been fully emphasized yet. As the containment measures during an epidemic can have severe consequences (or costs) on a region's society and economy (though they may effectively hinder the spread of diseases)\cite{Acemoglu:2021,Farboodi:2021}, it is of paramount importance to address the questions such as when and how to restore the social activity so that the costs can be minimized. To solve these problems, we need a comprehensive understanding of how the reversible change in contact networks can influence epidemic dynamics. 

Here, we analyze the conventional Susceptible-Infected-Removed (SIR) model \cite{Satorras:2015,Ruan:2020} on a particular type of time-varying network which presents a reversible change in its structure. Specifically, the contact network would first freeze, with links becoming inactive with rate $\gamma$. After a time $t^*$, the frozen links begin to recover and become active again with rate $\eta$. We show that all the three parameters $\gamma$, $\eta$ and $t^*$ are significant in the spreading dynamics. In particular, we find that the recovery of the network may markedly alter the epidemic shape, leading to the resurgence of the epidemic. This effect consequently brings about a counterintuitive result: there exists an optimal value of freezing rate which corresponds to the smallest epidemic size. We further extend this model (which is an isolated network) by letting the nodes interact weakly with the outside environment. The new model has the potential to explain the observational data which shows that in some regions, the epidemic could resurge even if the infected cases have dropped to zero. 


\section*{Epidemic spreading in an isolated network}

Let us begin by considering the SIR model defined on an isolated Erd\H{o}s-R\'enyi (ER) random network with average degree $\langle k\rangle$. Nodes in the network are divided into three compartments: infected (I), susceptible (S), and removed (R), corresponding to different states of individuals in the contagion process. Let $S_t$, $I_t$, and $R_t$ be the fraction of susceptible, infected, and removed nodes at time $t$, respectively. We have 
\begin{eqnarray}\label{eq:population}
S_t+I_t+R_t=1.
\end{eqnarray}
At each time step, every susceptible node is infected with probability $\beta$ if it is connected to one infected node (hence, the node is infected with probability $k\beta$ if it is linked to $k$ infected nodes). At the same time, each infected node becomes removed with probability $\mu$.  

We suppose the underlying contact network varies with time as follows: Initially, all links in the network are in an active state (meaning that a disease could normally transmit through these links). As the epidemic spreads, the system enters into a freezing stage, where each link freezes  (or we can say that these links become inactive) with probability $\gamma$ at each time step. Until time $t^*$, the system starts to recover, and the frozen links become active again with rate $\eta$. Denote $a_t$ as the fraction of active links at time $t$. The dynamics are governed by the following set of differential equations
\begin{eqnarray}\label{eq:dynamics}
\frac{dS_t}{dt}&=&-\beta \langle k\rangle a_t I_t S_t,  \label{eq:dynamics1}\\  
\frac{dI_t}{dt}&=&\beta \langle k\rangle a_t I_t S_t -\mu I_t, \label{eq:dynamics2}\\ 
\frac{dR_t}{dt}&=&\mu I_t,  \label{eq:dynamics3}
\end{eqnarray}
with

\begin{eqnarray}\label{eq:a_t)}
\frac{da_t}{dt}=
\begin{cases}
-\gamma a_t, ~~~ t\le t^* \\
\eta (1-a_t), ~~~ t>t^*. 
\end{cases}
\end{eqnarray}


Assuming that only a very small fraction of nodes are infected at the beginning, the initial conditions are $S_0 \approx 1$, $I_0 \approx 0$, $R_0 = 0$ and $a_0=1$. From Eq.(\ref{eq:a_t)}), we obtain that 
\begin{eqnarray}\label{eq:a_t2}
a_t=
\begin{cases}
e^{-\gamma t}, ~~~ t\le t^* \\
1-(1-e^{-\gamma t^*})e^{-\eta(t-t^*)}, ~~~ t>t^*.
\end{cases}
\end{eqnarray}

\begin{figure}
\epsfig{figure=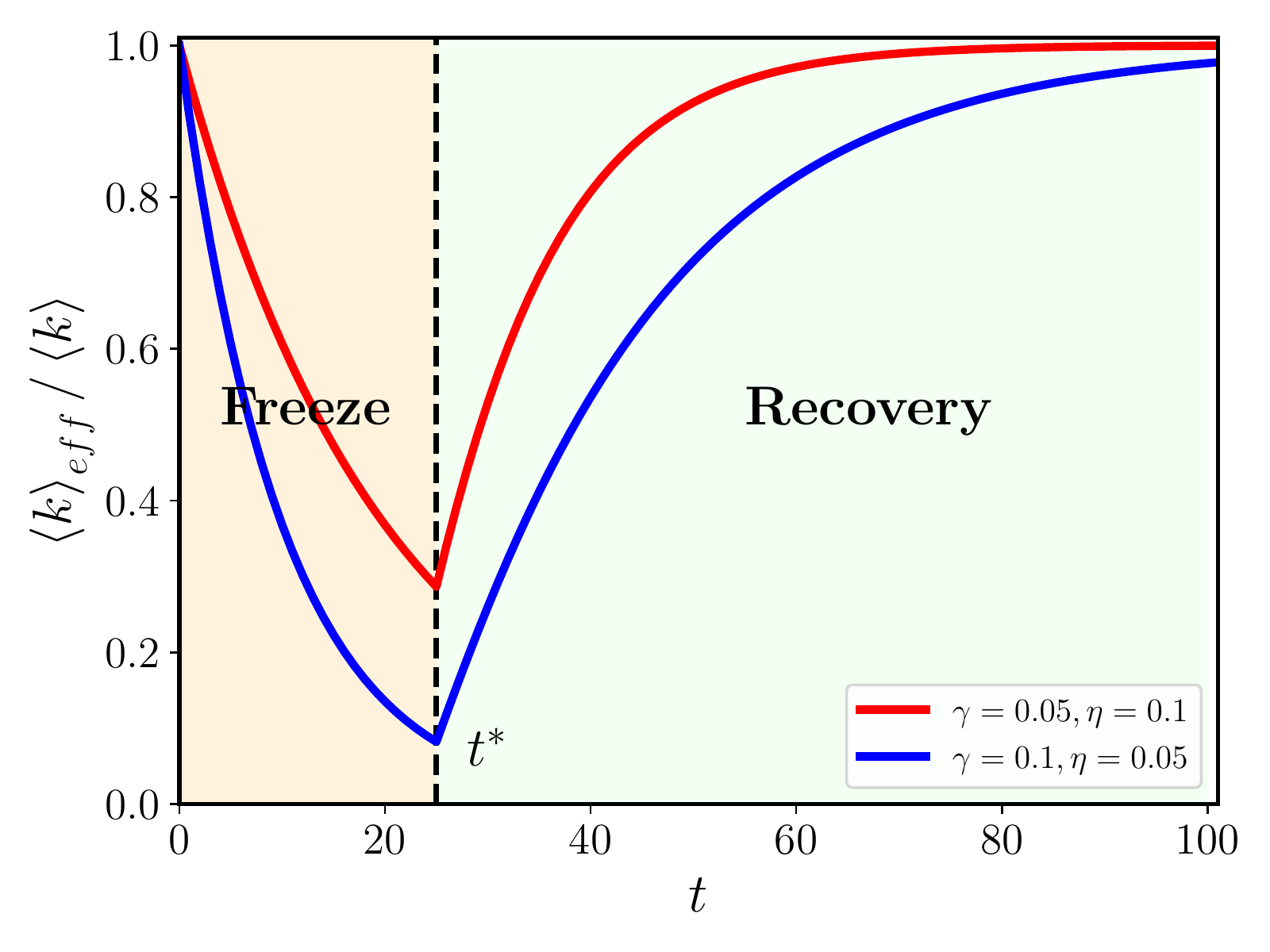,width=1.0\linewidth} \caption{(color online). Fraction of active links $\langle k\rangle_{eff}/\langle k\rangle$ as a function of time $t$ for different values of $\gamma$ and $\eta$. The spreading process is divided into two stages: For $t \le t^*$, the links freeze with rate $\gamma$, and the average contact number declines exponentially; while for $t>t^*$, the frozen links start to recover with rate $\eta$. In the schematic diagram, $t^*=25$ and $\langle k\rangle=8$.} \label{Fig:2}
\end{figure}

The average number of active links (or the average contact number for each node) is $\langle k\rangle_{eff} = \langle k\rangle a_t$. Figure \ref{Fig:2} illustrates $\langle k\rangle_{eff}$ as a function of time $t$ according to Eq.(\ref{eq:a_t2}). Evidently, the variation in $\langle k\rangle_{eff}$ qualitatively characterizes the change in the contact patterns of individuals in real life. Since the freezing rate $\gamma$, the recovery rate $\eta$, and the start time of recovery $t^*$ have direct effects on the contact patterns, we aim to understand how these parameters may further influence the epidemic dynamics.

At the very beginning of the epidemic (in the freezing stage), Eq.(\ref{eq:dynamics2}) can be linearized as
\begin{eqnarray}\label{eq:evolution_infected} 
\frac{dI_t}{dt}&=&\beta \langle k\rangle e^{-\gamma t} I_t  -\mu I_t \label{eq:linear},
\end{eqnarray}
where the terms of higher orders such as $I_t^2$ and $I_tR_t$ have been neglected. Solving Eq. (\ref{eq:evolution_infected}), we obtain
\begin{eqnarray}\label{eq:evolution_infected_solution} 
I_t=I_0e^{\Delta(t)},
\end{eqnarray}
with
\begin{eqnarray}\label{eq:Delta} 
\Delta(t)=\beta \langle k\rangle \frac{1-e^{-\gamma t}}{\gamma}-\mu t.
\end{eqnarray}
Note that when $\gamma \to 0$, $\Delta(t) \to (\beta \langle k\rangle-\mu)t$. In other words, $I_t\sim e^{(\beta \langle k\rangle-\mu)t}$, which turns back to the result of the standard SIR model on random ER networks. In this case, a small fraction of infecteds could yield an exponential growth in $I_t$ at initial times if $\frac{\beta}{\mu}>\frac{1}{\langle k\rangle}$. However, by expanding the right-hand side of Eq.(\ref{eq:Delta}) [$\Delta(t)=(\beta \langle k\rangle -\mu) t-O(\gamma^2 t^2)$ for small values of $\gamma t$], we find that in our model the growth in infected nodes is subexponential (slower than the exponential growth). This result may provide a new explanation for the observed empirical data of the early confirmed COVID-19 cases in China \cite{Maier:2020}.

\begin{figure}
\epsfig{figure=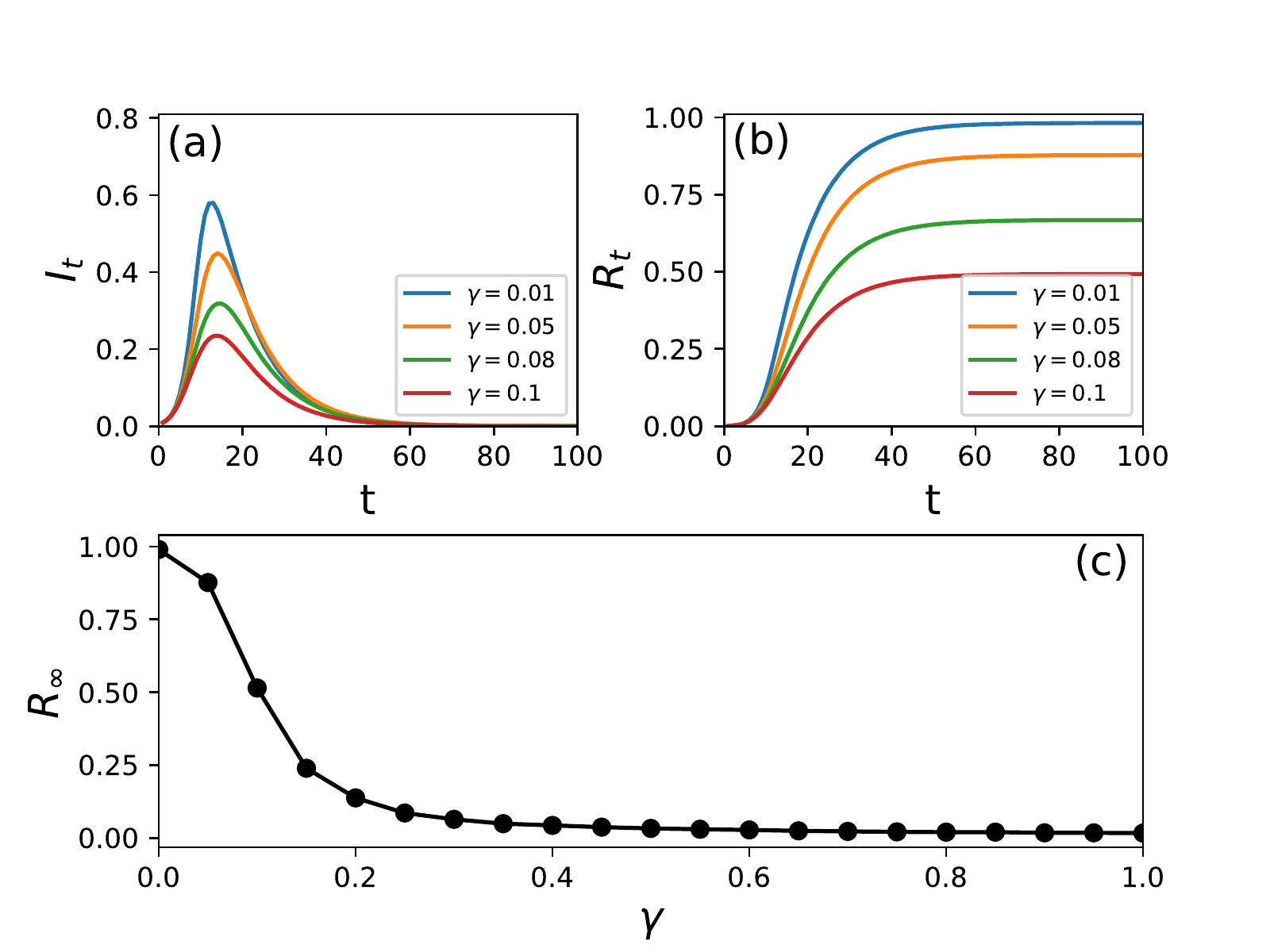,width=1.0\linewidth} \caption{(color online). Fraction of (a) infected  and (b) removed nodes as a function of time $t$ for different values of $\gamma$ without the recovery process, i.e., $\eta=0$. (c) Final fraction of removed nodes as a function of $\gamma$. The parameters in simulations are given as: $N=5000$, $\langle k\rangle=8$, $\beta=0.12$, $\mu=0.1$, and $t^*=25$. All results are averaged over $100$ realizations.} \label{Fig:3}
\end{figure}

To investigate in more detail the influence of $\gamma$ on epidemic dynamics, we first fix the recovery rate $\eta=0$. Figure \ref{Fig:3} (a) and (b) show the fraction of infected and removed nodes as a function of time $t$ for different values of $\gamma$, respectively. We see that increasing the freezing rate can significantly suppress the epidemic. Specifically, the shape of the $I_t-t$ curve becomes flattened [Fig.\ref{Fig:3}(a)], and the asymptotic value of $R_t$ decreases [Fig.\ref{Fig:3}(b) and (c)]. In the freezing stage, the average contact number $\langle k\rangle_{eff}$ of each node decreases exponentially as $t$ increases. When $\langle k\rangle_{eff}$ drops below the critical value $1$, the underlying ER network falls into small pieces; correspondingly, the spread almost stops. As a consequence, for large $\gamma$, the disease can only spread a few steps due to the rapid fragmentation of the network, resulting in a very small number of nodes being infected eventually [Fig.\ref{Fig:3}(c)]. 


\begin{figure}
\epsfig{figure=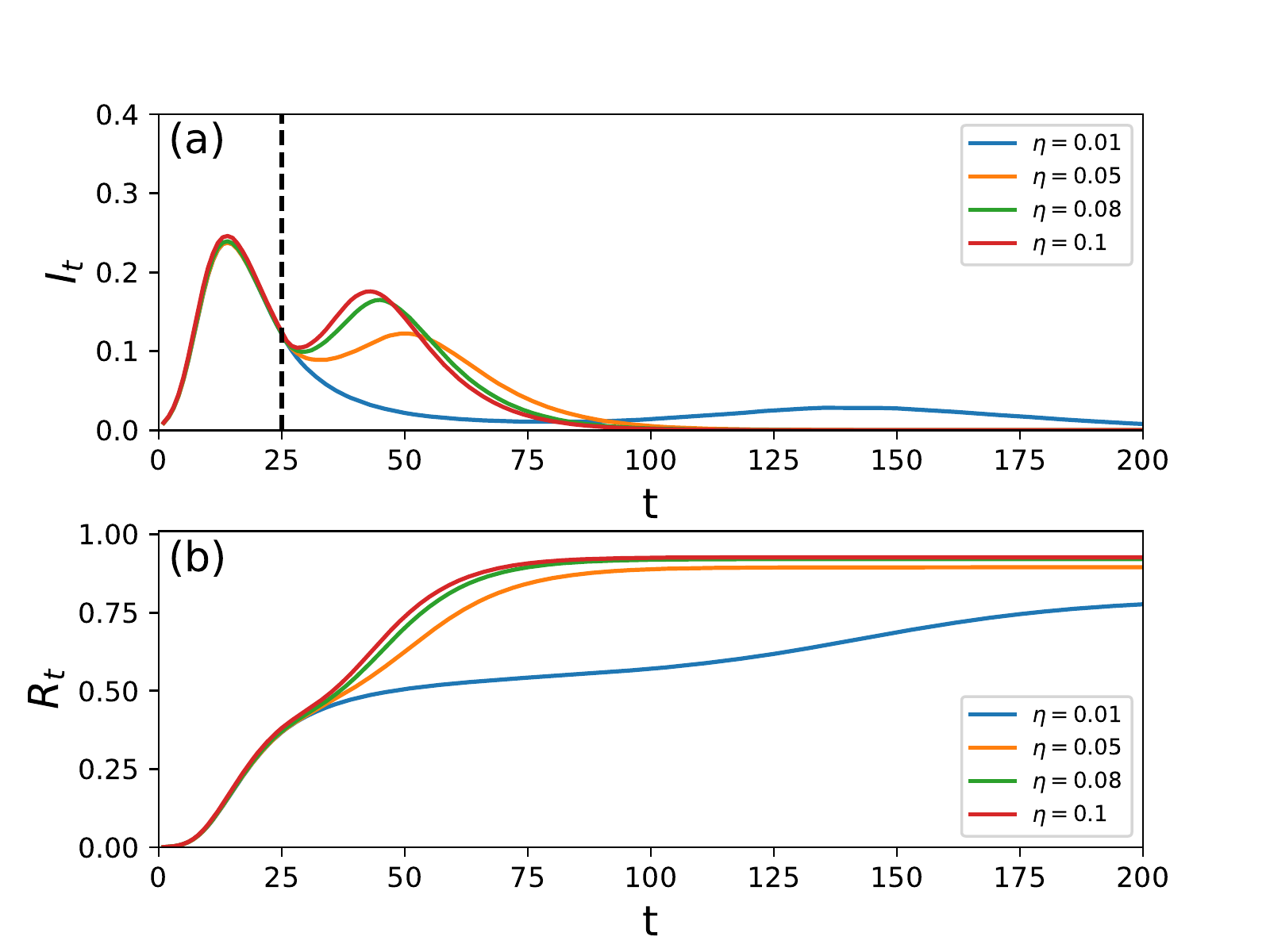,width=1.0\linewidth} \caption{(color online).  Fraction of (a) infected  and (b) removed nodes as a function of time $t$ for different values of $\eta$. The parameters in simulations are given as: $N=5000$, $\langle k\rangle=8$, $\beta=0.12$, $\mu=0.1$, $\gamma=0.01$, and $t^*=25$. All results are averaged over $100$ realizations.} \label{Fig:4}
\end{figure}

Introducing link recovery could prominently change the epidemic shape. As shown in Fig.\ref{Fig:4} (a), we observe that increasing $\eta$ may lead to the resurgence of the epidemic --- a second wave appears\cite{Aleta:2020}. A closer inspection shows that the peak value of the second wave rises as $\eta$ increases, while the time to reach the peak reduces, implying that the rapid recovery of frozen links (for large values of $\eta$) may exacerbate the spreading. Such effect is further confirmed by Fig.\ref{Fig:4} (b), where we show how the fraction of removed nodes varies with time for different values of $\eta$. It is worth noticing that the turning point (the local minimum between the two peaks)  in the $I_t-t$ curve occurs after $t^*=25$ --- the critical transition time of the contact network. This delay comes about because the system has to take some time to respond to the abrupt change in the structure. To better understand, notice that only if enough frozen links have recovered (which takes a while) such that the infection process ($S \to I$) outcompetes the removal process ($I \to R$) again, a rebound in the declining phase of the epidemic is possible.


\begin{figure}
\epsfig{figure=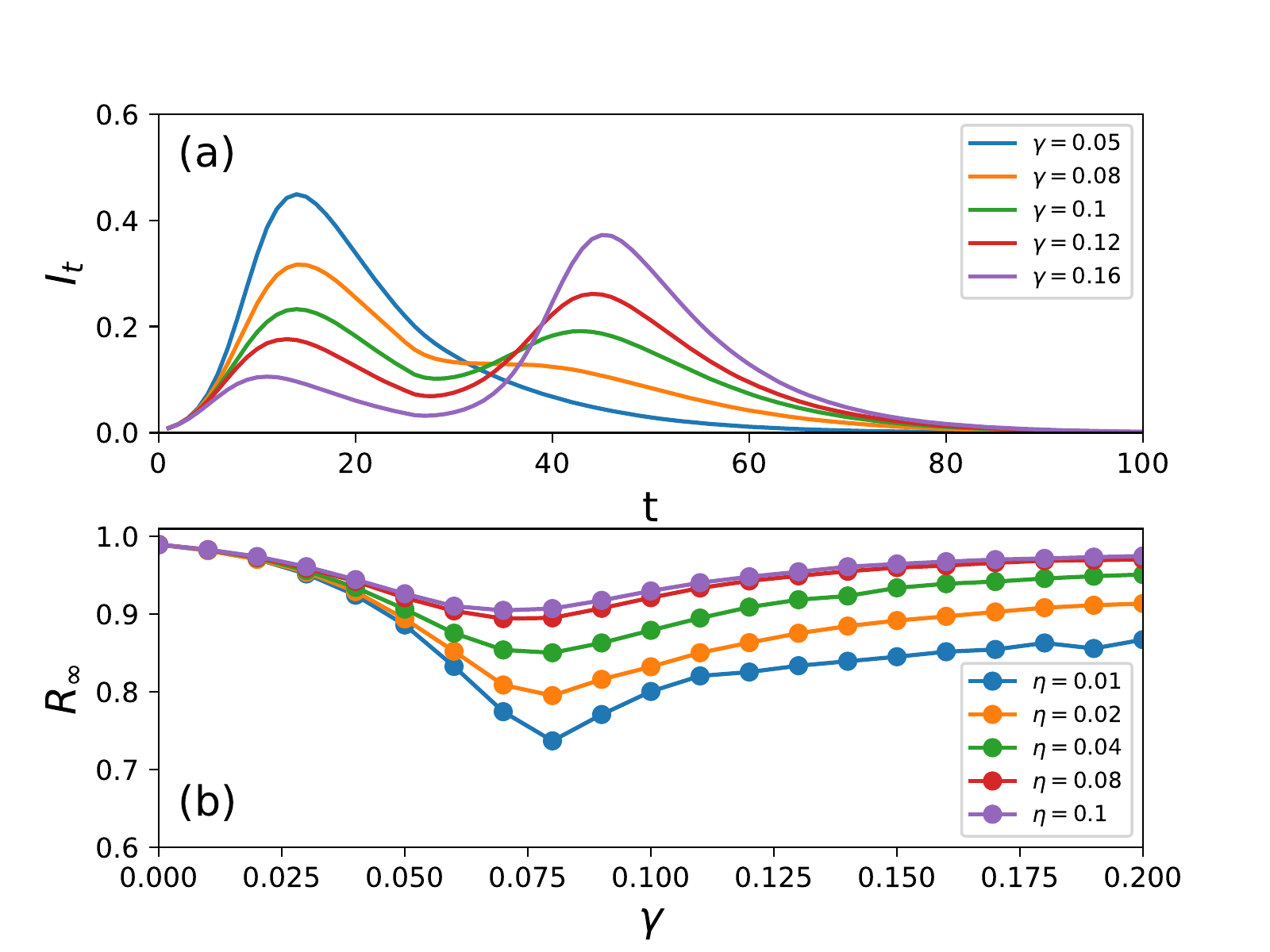,width=1.0\linewidth} \caption{(color online). (a) Fraction of infected nodes changes with time $t$ for different values of $\gamma$ given $\eta=0.1$. (b) Final fraction of removed nodes changes with $\gamma$ for different values of $\eta$. The parameters in simulations are given as: $N=5000$, $\langle k\rangle=8$, $\beta=0.12$, $\mu=0.1$, and $t^*=25$. All results are averaged over $100$ realizations.} \label{Fig:5}
\end{figure}

Further analysis shows that the joint action of the freeze and recovery processes could lead to some interesting phenomena. When $\gamma$ is small, the epidemic can not be effectively contained at the freezing stage, which results in a high peak in the infection evolution curve. In this case, the susceptible nodes are largely consumed and can not provide enough fuels for the second wave, making it incidental. While for large $\gamma$, there will be a large enough pool of susceptible nodes, and if the number of infected nodes is not too small, a stronger bounce at the second stage becomes possible [see Fig.\ref{Fig:5}(a)]. The transformation in the epidemic shape hints that the final epidemic size may change with the freezing rate in a nontrivial way. Figure \ref{Fig:5}(b) manifests how the final fraction of removed nodes varies with $\gamma$ for different values of $\eta$. We see that given $\eta>0$ there is an optimal value of $\gamma$ which corresponds to the lowest epidemic prevalence level. Surprisingly, we find that the optimal value is not sensitive to $\eta$, which keeps at around $\gamma \approx 0.08$.


\begin{figure}
\epsfig{figure=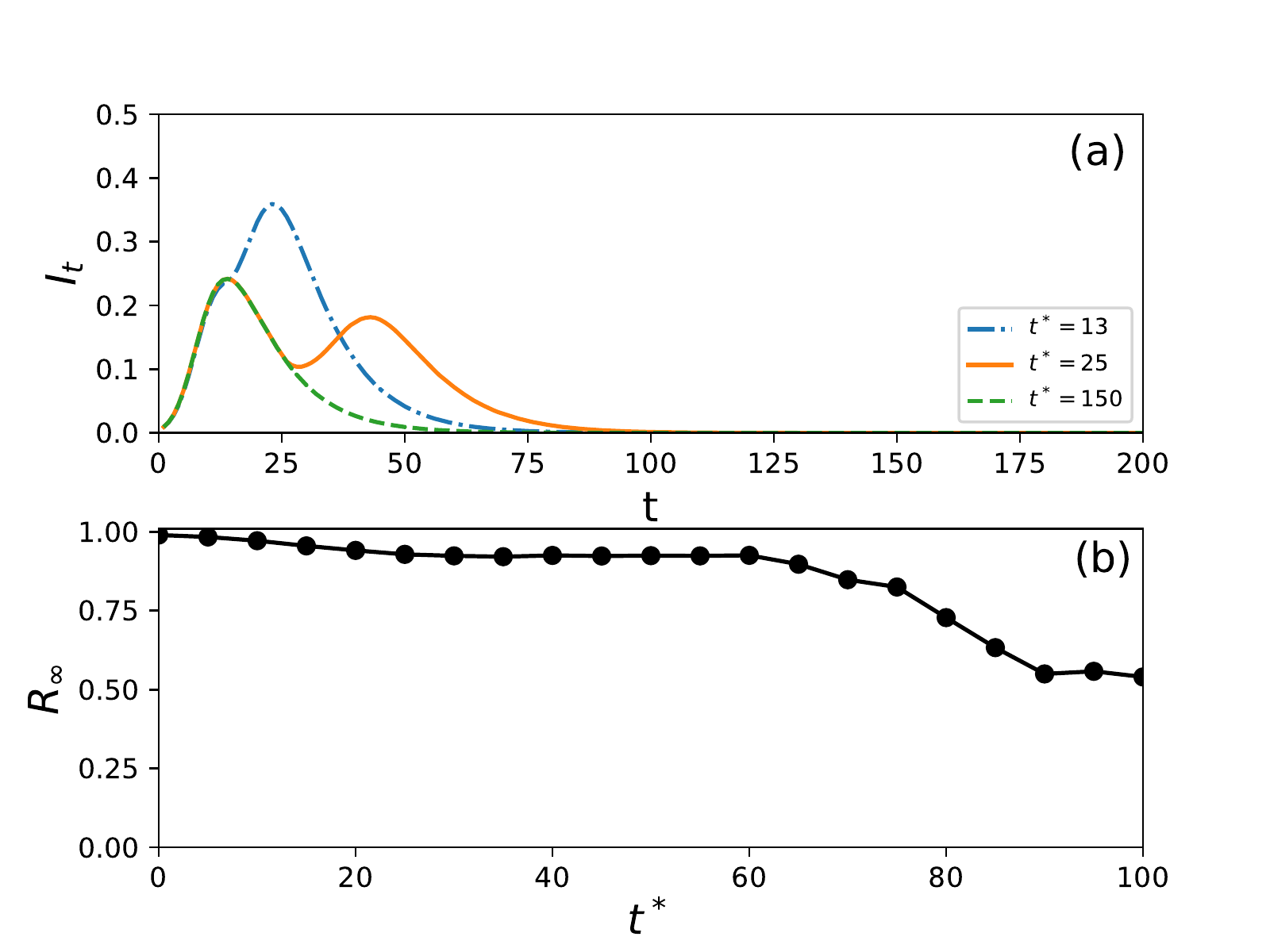,width=1.0\linewidth} \caption{(color online). (a) Fraction of infected nodes as a function of time $t$ for different values of $t^*$. (b) Final fraction of removed nodes as a function of $t^*$. The parameters in simulations are given as: $N=5000$, $\langle k\rangle=8$, $\beta=0.12$, $\mu=0.1$, $\gamma=0.1$, and $\eta=0.1$. All results are averaged over $100$ realizations.} \label{Fig:6}
\end{figure}

The start time of recovery $t^*$ also plays a critical role in the spreading dynamics. Figure \ref{Fig:6}(a) demonstrates that only for intermediate values of $t^*$ one can observe two peaks in the infection evolution curve. For small values of $t^*$, the epidemic is still in growing stage. In this case, the recovery of links may accelerate the spreading process, making the infected number increase to a higher peak than expected in the case of no recovery process, i.e., $\eta=0$. After then, it declines monotonously.  While for large $t^*$ (when the spreading dynamics has ended in the case of $\eta=0$), no more infected nodes remain, and the resurgence of the epidemic becomes impossible. Figure \ref{Fig:6}(b) shows the final fraction of removed nodes as a function of $t^*$. We see that as $t^*$ increases, $R_\infty$ is suppressed, which leads to an intuitive conclusion that postponing the start time of recovery can help to mitigate the spread of an epidemic.

\section*{Epidemic spreading in an open network}

\begin{figure}
\epsfig{figure=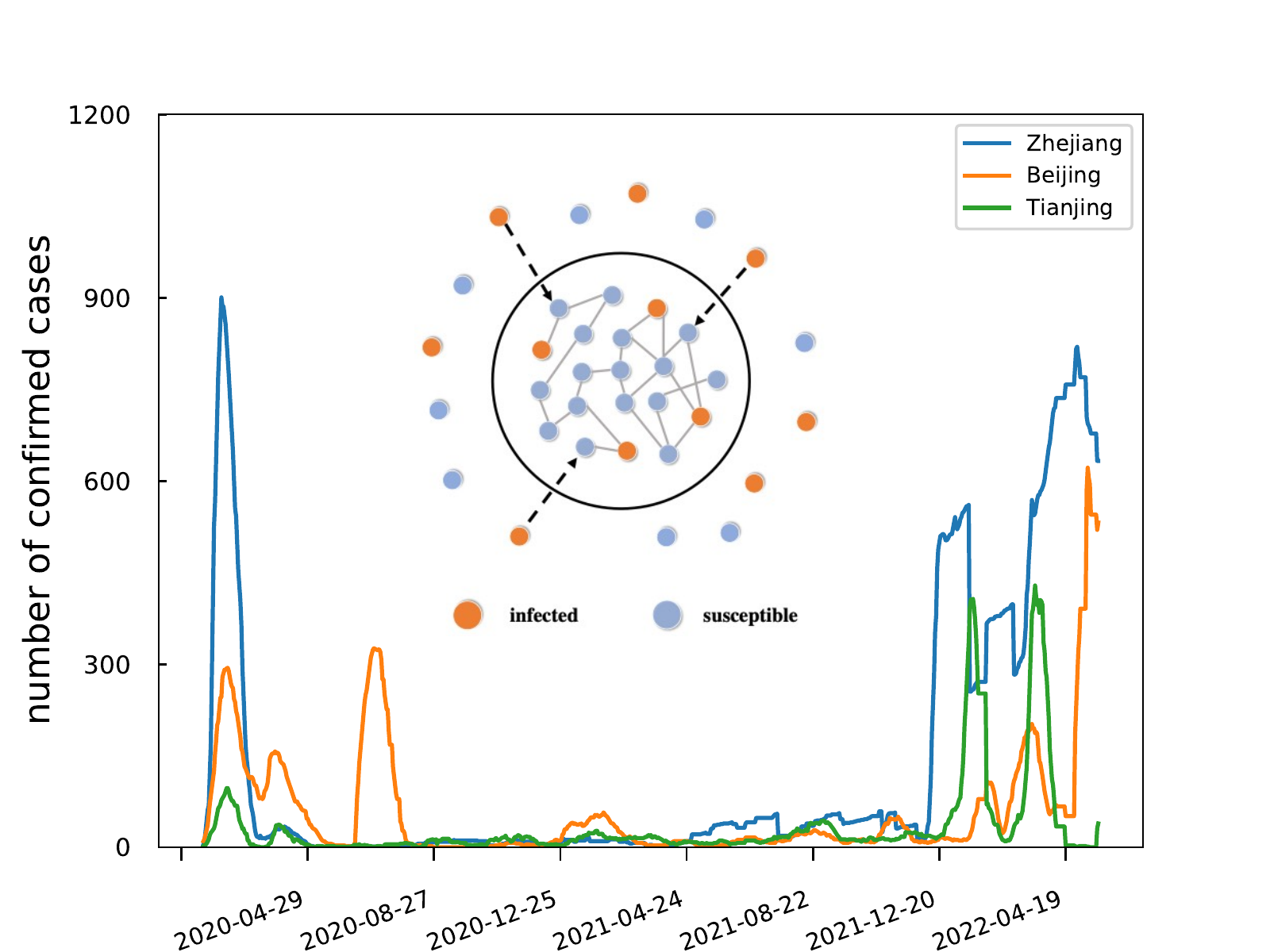,width=1.0\linewidth} \caption{(color online). Case numbers in three different districts in Mainland China. These regions share the same phenomenon: the epidemic resurges after the number of confirmed cases has declined to zero (lasting for a period of time). The inset graph shows an open network interacts weakly with its environment. We assume a small fraction of susceptibles inside the system may acquire the disease from the outside infecteds due to the recurrent movements of the individuals. }\label{Fig:7}
\end{figure}

The above model can partly illuminate some phenomena observed in the COVID-19 pandemic, such as double-waves and subexponential growth in infecteds at early times. Nevertheless, it fails to explain the following situation: the epidemic may resurge even if it has died out in a region (see Fig.\ref{Fig:7}, where the infection data are collected from Ding Xiang Yuan $\footnote{https://ncov.dxy.cn/}$). The main limitation of the previous model is that it is isolated. While in reality, individuals may acquire the disease from the outside agents (not in the current region, since individuals would travel recurrently between different regions \cite{Belik:2011,Ruan:2015,Wang:2013,Ruan:2017,Poletto:2012}). To mimic this process, we modify the previous model as follows

\begin{eqnarray}\label{eq:dynamics}
\frac{dS_t}{dt}&=&-\beta \langle k\rangle a_t I_t S_t -qS_t, \label{eq:dynamics2-1}\\  
\frac{dI_t}{dt}&=&\beta \langle k\rangle a_t I_t S_t -\mu I_t +qS_t,\label{eq:dynamics2-2}\\ 
\frac{dR_t}{dt}&=&\mu I_t,  \label{eq:dynamics2-3}
\end{eqnarray}
with
\begin{eqnarray}\label{eq:a_t-2)}
\frac{da_t}{dt}=
\begin{cases}
-\gamma a_t, ~~~ t\le t^* \\
\eta (1-a_t), ~~~ t>t^*.
\end{cases}
\end{eqnarray}
Here, we have assumed that a small fraction $q \ll 1$ of susceptibles may acquire the disease from the ``environment" at each time step (or we can say that the susceptible nodes become infected ``spontaneously" with rate $q$ \cite{Zheng:2015}). This model is crude since it ignores the exchange in nodes between the system and its environment. If we consider the open system is in an equilibrium state, i.e., the number of nodes leaving (at random) from the contact network (with each node taking away $\langle k\rangle$ links) at each time step equals to the number of new nodes (with each node establishing $\langle k\rangle$ new links to other randomly chosen nodes) joining in, the underlying network is statistically unchanged in the structure. From this point of view, our model is reasonable.  

For extremely small values of $q$, the effect of the ``external" term $qS_t$ on epidemic dynamics is subtle if the recovery process is not taken into account. Note that $q$ is a slow variable in the differential equations, which at first glance gives the impression that both $S_t$ and $I_t$ would change slowly for large $t>T_c$, where $T_c$ is defined as the time to reach the steady state in the case of $q=0$ (before $T_c$ the dynamical equations are governed by the terms $\beta \langle k\rangle a_t I_t S_t$ and $\mu I_t$). However, since $q \ll \mu$, the newly infected nodes induced by the environment (the contribution from the term $qS_t$) would immediately turn into the removed state. As a result, instead of $I_t$, $R_t$ will increase slowly, while $I_t$ keeps unchanged for large $t$, which equals to $0$.

\begin{figure}
\epsfig{figure=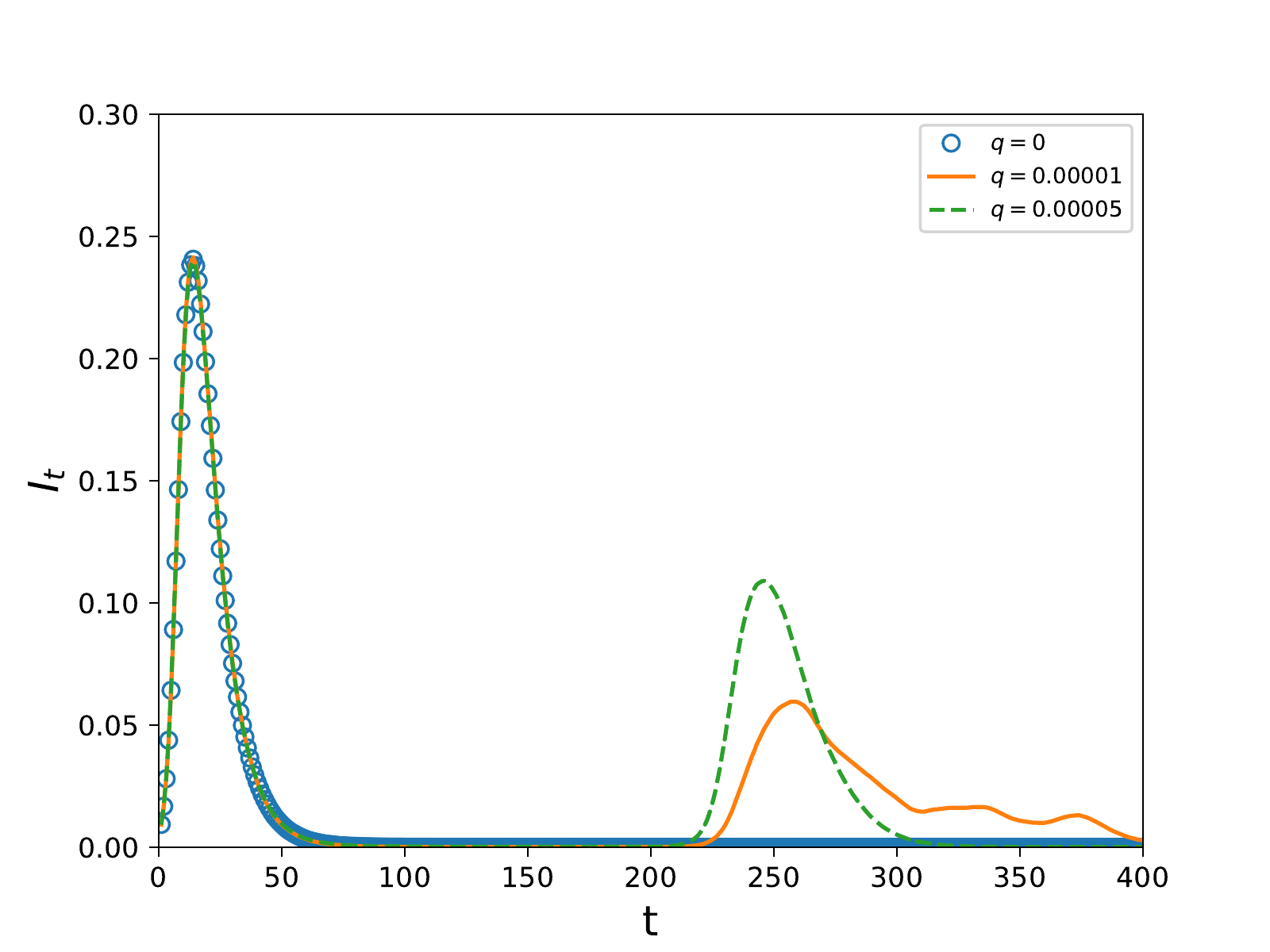,width=1.0\linewidth} \caption{(color online).  Fraction of infected nodes as a function of time $t$ for $q=0$, $10^{-5}$ and $5 \times 10^{-5}$. The other parameters in simulations are: $N=5000$, $\langle k\rangle=8$, $\beta=0.12$, $\mu=0.1$, $\gamma=0.1$, $\eta=0.1$, and $t^*=200$. All results are averaged over $100$ realizations.} \label{Fig:8}
\end{figure}

Incorporating the recovery mechanism could yield a notable change in the epidemic evolution process. In this case, the perturbation from the environment is amplified by link recovery, which may result in a fast increase in the infected number at the recovery stage (the contribution from the term $\beta \langle k\rangle a_t I_t S_t$).  In particular, when the increasing rate in the infecteds exceeds the removed rate, a second wave in the $I_t-t$ curve could appear. Figure \ref{Fig:8} confirms our analysis, where we see that even though the number of infected nodes in the contact network has declined to zero, the epidemic could resurge due to the synergetic action of external perturbations and internal link recovery. 

\section*{Discussion}

In this study, based on the real data of human mobility during the COVID-19 pandemic, we found that the mobility pattern exhibits a reversible change over time. We supposed that the underlying contact network of individuals would experience such a change as well. On the basis of this assumption, we investigated the classical SIR model on a time-varying network, where the links would undergo a freeze-recovery process. Specifically, we assumed there is a turning point $t^*$ in the timeline: before $t^*$ the links freeze with a certain rate $\gamma$; while after it, the frozen links recover with rate $\eta$. 

Our analysis shows that all the three parameters $\gamma$, $\eta$ and $t^*$ are important in the epidemic dynamics. We first focused on isolated networks and demonstrated that the freezing process may lead to the subexponential growth in the number of infected nodes at initial times. Its effect on the epidemic size, however, intriguingly depends on the recovery rate $\eta$. Without recovery ($\eta=0$), the final epidemic size decreases monotonously with $\gamma$ until near $0$. While for $\eta>0$, there exists an optimal value of $\gamma$ which corresponds to the smallest epidemic size. The reason lies in the recovery mechanism which could give rise to epidemic resurgence. Unexpectedly, we found that the optimal value is not sensitive to $\eta>0$. Considering that in reality the recovery rate is highly heterogeneous across different regions (due to the diversity in economy and population levels), our result provides meaningful insight into the epidemic control in practice. Finally, we showed that postponing the start time of recovery $t^*$ may effectively mitigate the spread of the epidemic. 

The above model suggests that link recovery could cause epidemic resurgence provided that a few infected nodes still remain in the system. This condition however conflicts with the real situations, where we observe that in some regions the epidemic could rebound even if the number of infected cases has declined to zero. To explain this phenomenon, we further modified the model, taking into account the weak interactions between the contact network and its environment. The new model shows that the external perturbations (from the environment) could be amplified by the recovery mechanism, which leads to the appearance of the second wave even if the system is in an absorbing state (no infected nodes exist).


Our model could be extended to more complex scenarios, for instance, considering the multiple freeze-recovery cycles in contact networks, which may generate multi-waves in the infection evolution curve \cite{Perra:2011,Zheng:2018,Weitz:2020}. Besides, the network structure may also have a significant effect on the spreading dynamics \cite{Pastor-Satorras:2001,Moreno:2003,Liu:2005,Xuan:2013}, which needs further more investigations.

\section*{Acknowledgments}
This work was partially supported by the Zhejiang Provincial Natural Science Foundation of China under Grants No. LY21F030017 and No. LR19F030001, by the National Natural Science Foundation of China under Grant No. 61973273, and by the Key R\&D Programs of Zhejiang under Grant No. 2022C01018.

\end{document}